\documentclass{iau}
\usepackage{graphicx}

\def\Mpc{$h^{-1}$~{\rm  Mpc}}

\title[Cosmic Web Paradigm] 
{Yakov Zeldovich and the Cosmic Web Paradigm}

\author[Jaan Einasto]   
{Jaan Einasto$^1$}

\affiliation{$^1$Tartu Observatory, \\ Observatooriumi 1, 61602
  T\~oravere, Estonia  \\ email: {\tt jaan.einasto@to.ee} }

\pubyear{2015}
\volume{308}  
\pagerange{xxx--yyy}
\jname{The Zeldovich Universe: Genesis and Growth of the Cosmic Web}
\editors{Rien van de Weygaert, Sergei Shandarin, Enn Saar \%  Jaan Einasto, eds.}
\begin{document}

\maketitle

\begin{abstract}

I discuss the formation of the modern cosmological paradigm.  In more
detail I  describe the early study of dark matter and cosmic web and
the role of Yakov Zeldovich in the formation of the present concepts
on these subjects.  

\end{abstract}

\section{Formation of the modern cosmological paradigm}

The modern classical cosmological paradigm was elaborated step by step
during the first part of the 20th Century. It was found that there
exist stellar systems outside our Milky Way --- external galaxies
(\cite{Opik:1922}, \cite{Hubble:1925ij}). Next it was found that
external galaxies are moving away from us, i.e. the Universe is
expanding (\cite{Hubble:1929th}).  On the basis of Einstein relativity
theory \cite{Friedman:1922fk} explained the expansion as a property of
the infinite universe. The speed of the expansion can be expressed in
terms of the Hubble constant, $H_0$. \cite{Sandage:1975nx} found a
value of about $H_0 = 50$~km~s$^{-1}$~Mpc$^{-1}$, whereas
\cite{de-Vaucouleurs:1978d} and \cite{van-den-Bergh:1972mz} preferred
a value around $H_0 = 100$~km~s$^{-1}$~Mpc$^{-1}$. Due to this
uncertainty the Hubble constant is often expressed in dimensionless
units $h$, defined as: $H_0 = 100~h$~km~s$^{-1}$~Mpc$^{-1}$.

Another basis of the classical cosmological paradigm is the
distribution of galaxies and clusters of galaxies.  A photographic
survey was made using the 48-inch Palomar Schmidt telescope.
\cite{Abell:1958} used the Palomar survey to compile a catalogue of
rich clusters of galaxies for the Northern sky; later the catalogue
was continued to the Southern sky (\cite{Abell:1989}).
\cite{Zwicky:1968} used this survey to compile for the Northern
hemisphere a catalogue of galaxies and clusters of galaxies.  The
galaxy catalogue is complete up to 15.5 photographic magnitude.  Both
authors noticed that galaxies and clusters of galaxies show a tendency of
clustering.

A deeper complete photographic survey of galaxies was made in the Lick
Observatory by \cite{Shane:1967}. The Lick counts as well as galaxy
and cluster catalogues by Zwicky and Abell were analysed by Jim
Peebles and collaborators to exclude count limit irregularities
(\cite{Soneira:1978fk}).  These data show the apparent (2-dimensional)
distribution of galaxies and clusters on the sky.  The basic
conclusion from these studies was that galaxies are hierarchically
clustered.  There exist clusters and superclusters of galaxies, but
most galaxies form a more-or-less randomly distributed population of
field galaxies.

The mean density due to galaxies was determined using the mean
luminosity density, and the mean mass-to-luminosity ratio ($M/L$) of
galaxies.  Estimates available in the 1950's indicated a low-density
Universe, $\Omega \approx 0.05$.

This complex of data formed the classical cosmological paradigm.
However, the theoretical explanation of the expanding Universe by the
Friedman model was mathematical, it did not consider physical processes
in the early Universe.  Thus in the early 1960's in several centres
theorists started to think on the physics of the early Universe. Most
important developments in this direction were made in Princeton by Jim
Peebles, and in Moscow by Yakov Zeldovich and their collaborators. 

One of the first step in the study of physical processes was the
elaboration of the hierarchical clustering scenario of galaxies by
\cite{Peebles:1970}, \cite{Peebles:1971}.  On the other hand, the
Moscow team developed the pancake model of structure formation by
\cite{Zeldovich:1970}, and the theory of the Sunyaev-Zeldovich effect
in the Cosmic Microwave Background radiation by \cite{Sunyaev:1969tl}.

To discuss new problems of cosmology and astrophysics Zeldovich
organised summer and winter schools. The first of such schools was in
the new observatory in T\~oravere, 1962; later schools were hold in
Caucasus winter resorts. Our Tartu cosmology team was invited to
Caucasus winter schools in 1972, 1974 and later.  Discussions on
winter schools started our collaboration with the Zeldovich team.

Also important observational discoveries were
made.  \cite{Penzias:1965} detected the Cosmic Microwave Background
radiation.  Satellite observatories allowed to detect X-rays from
clusters of galaxies, and to find the mass of the hot gas in clusters,
as well as the total mass of clusters (\cite{Forman:1972zr}). 

These observational and theoretical developments were the basis of the
formation of the modern cosmological paradigm. In the following I
shall discuss in more detail some aspects of the new cosmological
paradigm, related to the discovery of the cosmic web.  Quite unexpectedly
it was found that the structure of the cosmic web is closely related to
another problem --- the nature of the dark matter and its role in the
formation of the cosmic web.

\section{Dark matter}

In the middle of 1960's the general opinion of the astronomical
community was that the classical cosmological paradigm is in agreement
with all observational and theoretical data available.  
Actually there were some unexplained facts. One of these curious data
was the Coma cluster mass paradox. The mass calculated from random
motions of galaxies in the cluster was much higher than the expected
mass found by adding masses of individual galaxies, as suggested by
\cite{Zwicky:1933}.

Another curious fact was the form of rotation curves of galaxies.  As
found by \cite{Oort:1940}, \cite{Roberts:1966dt} and
\cite{Rubin:1970},  the rotation curves of spiral galaxies are flat on
large galactocentric distances.  Since the surface brightness of
galaxies falls rapidly on the periphery, flat rotation curves mean,
that the mass-to-luminosity ratio rapidly increases on large
galactocentric distances. \cite{Oort:1940} and \cite{Roberts:1966dt}
explained this observation with the assumption that on large distances
low-mass stars dominate in galaxies. 

For some unclear reason these observations were ignored by the
astronomical community. 

Tartu astronomers have studied methods of modelling the structure of
galaxies for years. The first dynamical model of the Andromeda galaxy
was calculated by \cite{Opik:1922}. \cite{Kuzmin:1952a,Kuzmin:1956}
developed more accurate method of modelling galaxies, and applied the
method to our Galaxy.  

I helped Kuzmin in calculations and was interested to continue the
modelling of galaxies, using more observational data on galactic
populations.  First I studied carefully methods used by previous
authors to calculate mass distribution models of galaxies.  To my
surprise I found that in most models simple conditions of physical
reasonability are not satisfied.  Most important conditions are: the
spatial density must be non-negative and finite, some moments of the
density must be finite, in particular moments which define the mass
and the effective radius of the model.  Thus I found that the only density 
distribution profile which satisfies all physical conditions is a generalised
exponential model: $\varrho(a) = \varrho_0
\exp\left(-(a/a_c)^{1/N}\right)$, where $\varrho_0$ is the central
density, $a$ is the semi-major axis of the equidensity ellipsoid,
$a_c$ is the core radius, and $N$ is the structural parameter, which
allows one to vary the shape of the density profile. I used this
density profile in my model of the Galaxy (\cite{Einasto:1965}), and
in models of other galaxies.  Presently this profile is known as the
``Einasto profile''. 

The central problem in modelling galaxies is the calibration of
mass-to-luminosity ratios  of populations.  This can be done
using additional independent data. Most important data are velocity
dispersions of open and globular clusters with similar photometric
properties, assuming that galactic populations have been formed by
dissolution of clusters and star associations.  To bring data on
populations of different age and composition to a coherent system I
developed models of evolution of populations, similar to models by
\cite{Tinsley:1968}.  To my surprise I discovered that it is
impossible to represent rotation curves of galaxies by the sum of
gravitational attraction of known stellar populations.  The only way
to bring kinematical and photometrical data into agreement was to
suppose the presence of a new population --- corona --- with large
radius, mass and $M/L$ ratio.

I calculated models with massive coronas for all major galaxies of the
Local Group and the Virgo cluster central galaxy, M87. Results were
discussed at the First European Astronomy Meeting in Athens, September
1972 (\cite{Einasto:1974a}).  However, observed rotation curves were
not long enough to find the mass and the radius of coronas. Thus I
continued to think how to find total masses and radii of coronas.
Finally I decided to use companion galaxies as mass tracers of giant
galaxies. I collected data needed and found that the mass (and the
effective radius) of coronas is about ten times larger than the sum of
masses of all known stellar populations. The total cosmological
density of matter in galaxies including massive coronas is 0.2 of
critical cosmological density (\cite{Einasto:1974}).  A similar total 
density estimate was obtained by \cite{Ostriker:1974}.

I reported these results in the Arkh\~oz Winter School in January
1974. My principal conclusion was, that all giant galaxies have
massive coronas, and that coronas cannot have stellar nature. Thus the
coronal or dark matter is the principal constituent of the Universe,
and its nature is not clear.  After my talk Zeldovich invited me to
his room and asked two questions: Can we find data which give some
hints to the physical nature of coronas?  Can we find observational
evidence which can be used to discriminate between various theories of
the formation of galaxies?

To discuss the existence and the physical nature of dark matter, we
organised in January 1975 a conference in Tallinn, Estonia
(\cite{Doroshkevich:1975}).  The rumour on dark matter had spread
around the astronomical and physics community, and all leading Soviet
astronomers and physicists attended.  Two basic models were suggested
for coronae: faint stars or hot gas.  It was found that both models
have serious difficulties.  Neutrinos were also discussed but excluded
since they can form only clumps of rich cluster mass, but coronas of
galaxies have thousand times lower masses.

The dark matter problem was discussed also in the Third European
Astronomical Meeting in Tbilisi in June 1975. In the dark matter
session the principal discussion was between  supporters of the
classical paradigm with conventional mass estimates of galaxies, and
supportes of the new paradigm with dark matter. The most serious
arguments in favour of the classical cosmological paradigm were
presented by \cite{Materne:1976}: primordial nucleosynthesis suggests a
low-density Universe with  density parameter $\Omega \approx 0.05$
(this difficulty was already discussed by Zeldovich in the Tallinn
conference); the smoothness of the Hubble flow also favours a
low-density Universe. It was clear, that the existence of dark matter
is in conflict with the classical cosmological paradigm. If it exists,
then the density $\Omega \approx 0.2$ must be explained in some other
way.

The nature of dark matter and its role in the evolution of the
Universe was a problem for almost ten years.  To solve it data on the
distribution of galaxies in space and other new data were needed.

\section{Structure of the Universe}

When Zeldovich asked the question on the formation of galaxies I had
initially no idea how we could find an answer.  But soon I remembered
our previous experience in the study of galactic populations: their
spatial distribution and kinematics evolve slowly.  Systems of
galaxies are much larger in size, thus their evolution must be even
slower.  Random velocities of galaxies are of the order of several
hundred km/s or less, thus during the whole lifetime of the Universe
galaxies have moved from their place of origin only about 1~\Mpc.  If
there exist some regularities in the large-scale distribution of
galaxies, these regularities must reflect the conditions in the
Universe during the formation of galaxies. Thus we had a leading idea
to answer the Zeldovich question: {\em We have to study the
distribution of galaxies on large scales}.

We started to collect redshift data from all available sources. Since we
needed data on large-scale distribution of galaxies, we collected
redshifts not only for galaxies, but also for near cluster, both Abell
and Zwicky clusters, as well as active galaxies (radio and Markarian
galaxies). Our experience showed that clusters and active galaxies are
good tracers of the skeleton of the structure.  Redshifts of nearby
galaxies and clusters were searched for the whole Northern Hemisphere.

\begin{figure}[ht]
\centering
\resizebox{0.5\textwidth}{!}{\includegraphics{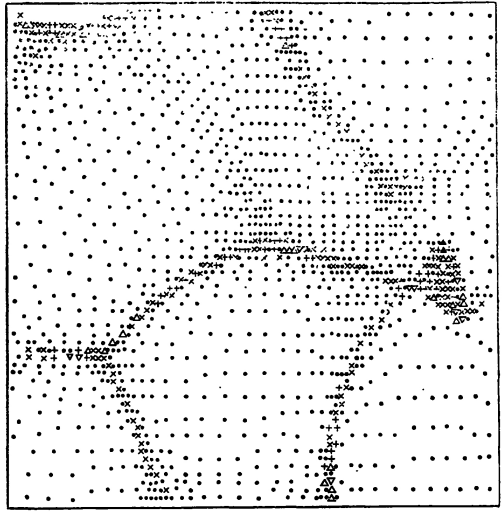}}
\caption{Distribution of particles in simulations according to
  Zeldovich pancake model (cited by \cite{Einasto:1980}).}
\label{fig:model}
\end{figure}

In the middle of 1970's there were two basic structure formation
scenarios, the \cite{Peebles:1970} hierarchical clustering scenario
and the \cite{Zeldovich:1970} pancake scenario.  The hierarchical
scenario represents well the apparent 2-dimensional distribution of
galaxies, seen in Lick maps.  Numerical experiments done in the
Zeldovich team showed the formation of high-density knots joined by
chains of particles to a connected network, as seen in
Figure~\ref{fig:model}.  Our challenge was to find out whether the
real distribution of galaxies shows similarity with some of these   
theoretical pictures.

\begin{figure}[ht]
\centering
\resizebox{0.325\textwidth}{!}{\includegraphics*{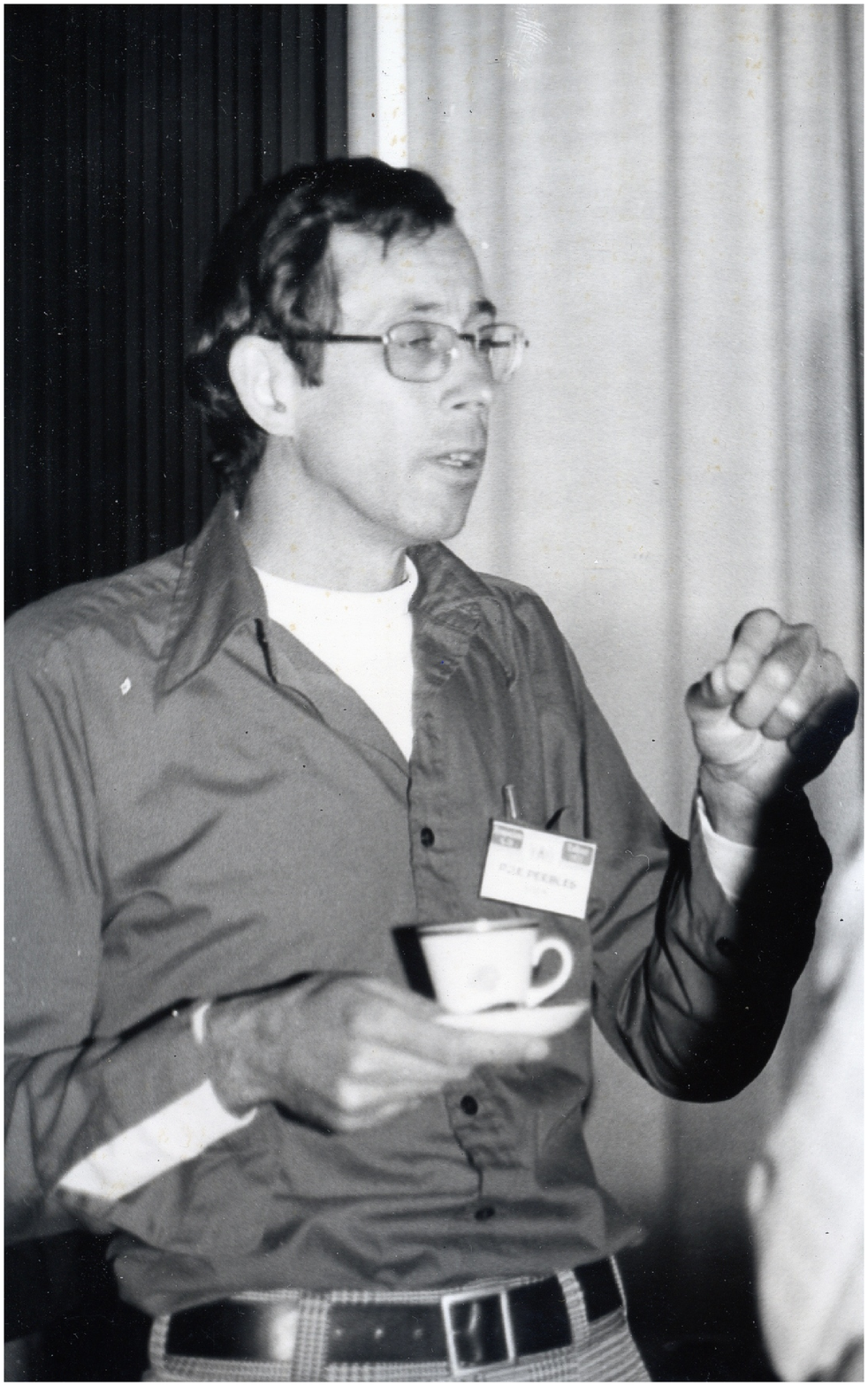}}
\resizebox{0.45\textwidth}{!}{\includegraphics*{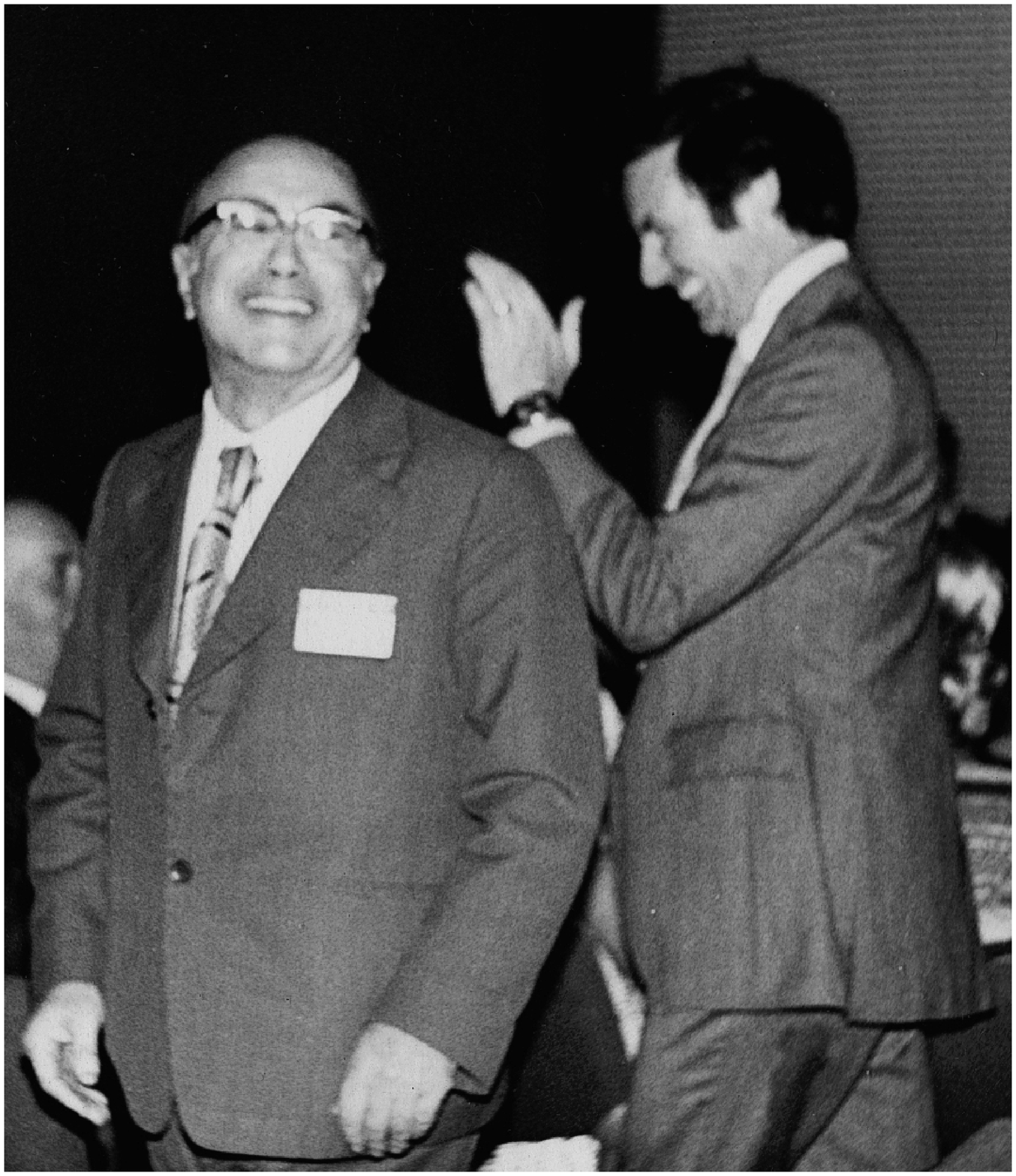}}\\
\caption{Jim Peebles (left), Yakov Zeldovich and Malcolm Longair
  (right) at the IAU Tallinn Symposium 1977 (author's archive).  }
\label{fig:tallinn77}
\end{figure}

After the Tbilisi Meeting Zeldovich proposed to organise an
international symposium devoted solely to cosmology.  This suggestion
was approved by IAU, and the symposium ``Large Scale Structure of the
Universe'' was hold in Tallinn in September 1977.  Two pictures of
participants are shown in Figure~\ref{fig:tallinn77}.

The first speaker on the distribution of galaxies was Brent Tully
(\cite{Tully:1978}), who showed a movie on the distribution of galaxies
in the Local Supercluster. The movie showed that the supercluster
consists of a number of galaxy chains which branch off from the
supercluster's central cluster. No galaxies could be seen in the space
between the chains.  The presence of voids in the distribution of
galaxies was reported also by \cite{Tifft:1978}, and
\cite{Tarenghi:1978} in the Coma and Hercules superclusters,
respectively.

\begin{figure}[ht]
\centering
\resizebox{0.40\textwidth}{!}{\includegraphics*{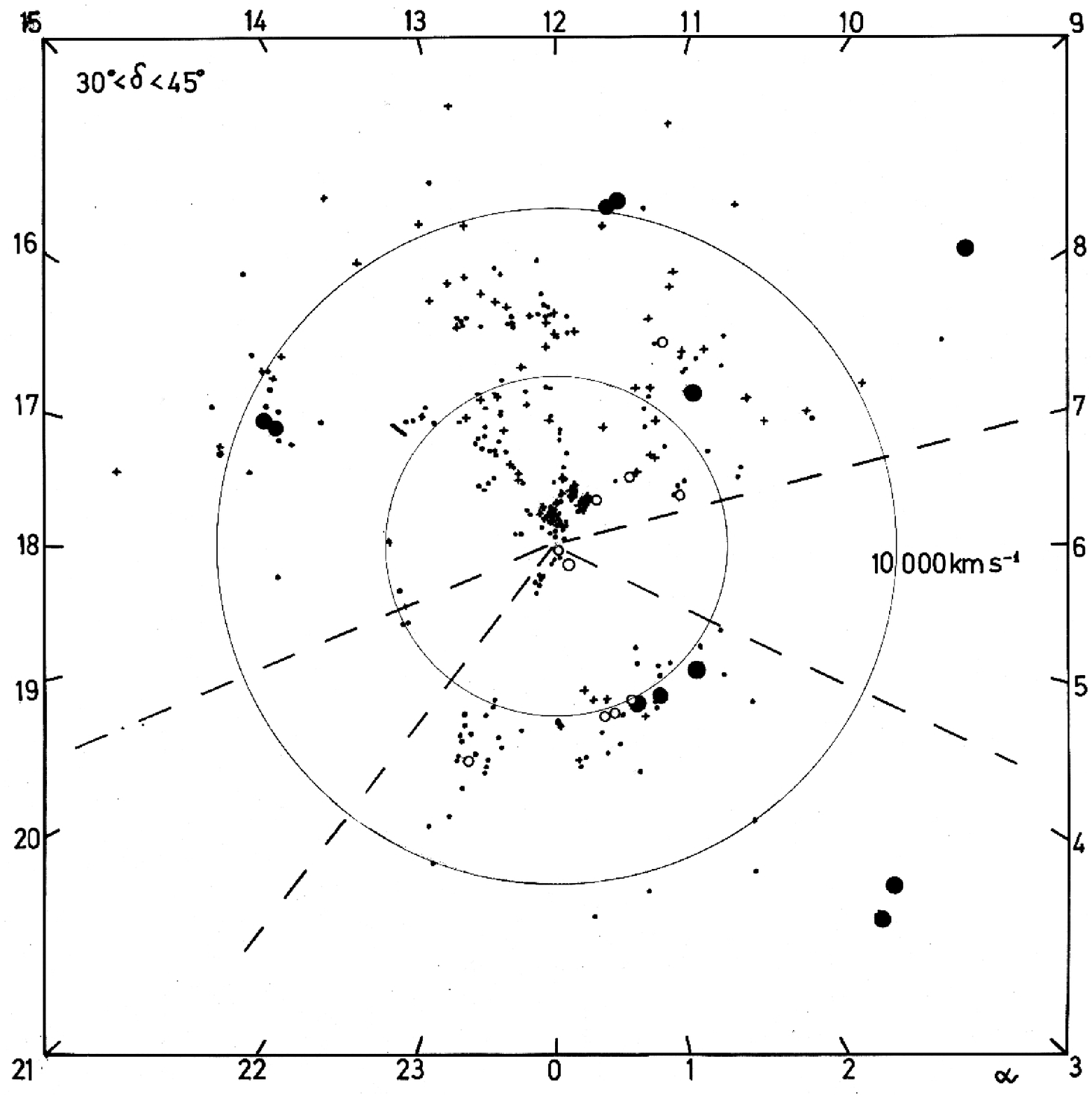}}
\resizebox{0.48\textwidth}{!}{\includegraphics*{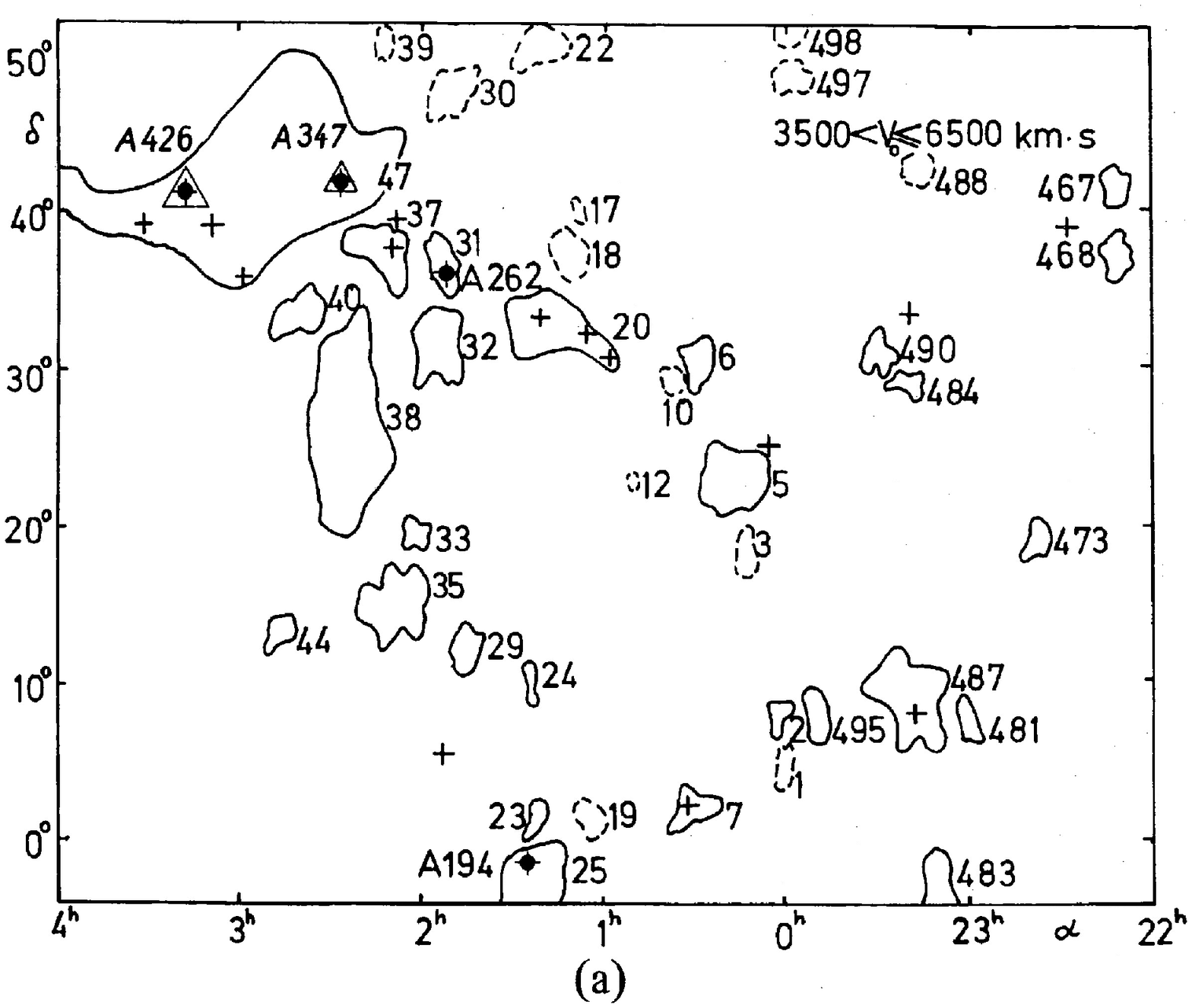}}\\
\resizebox{0.40\textwidth}{!}{\includegraphics*{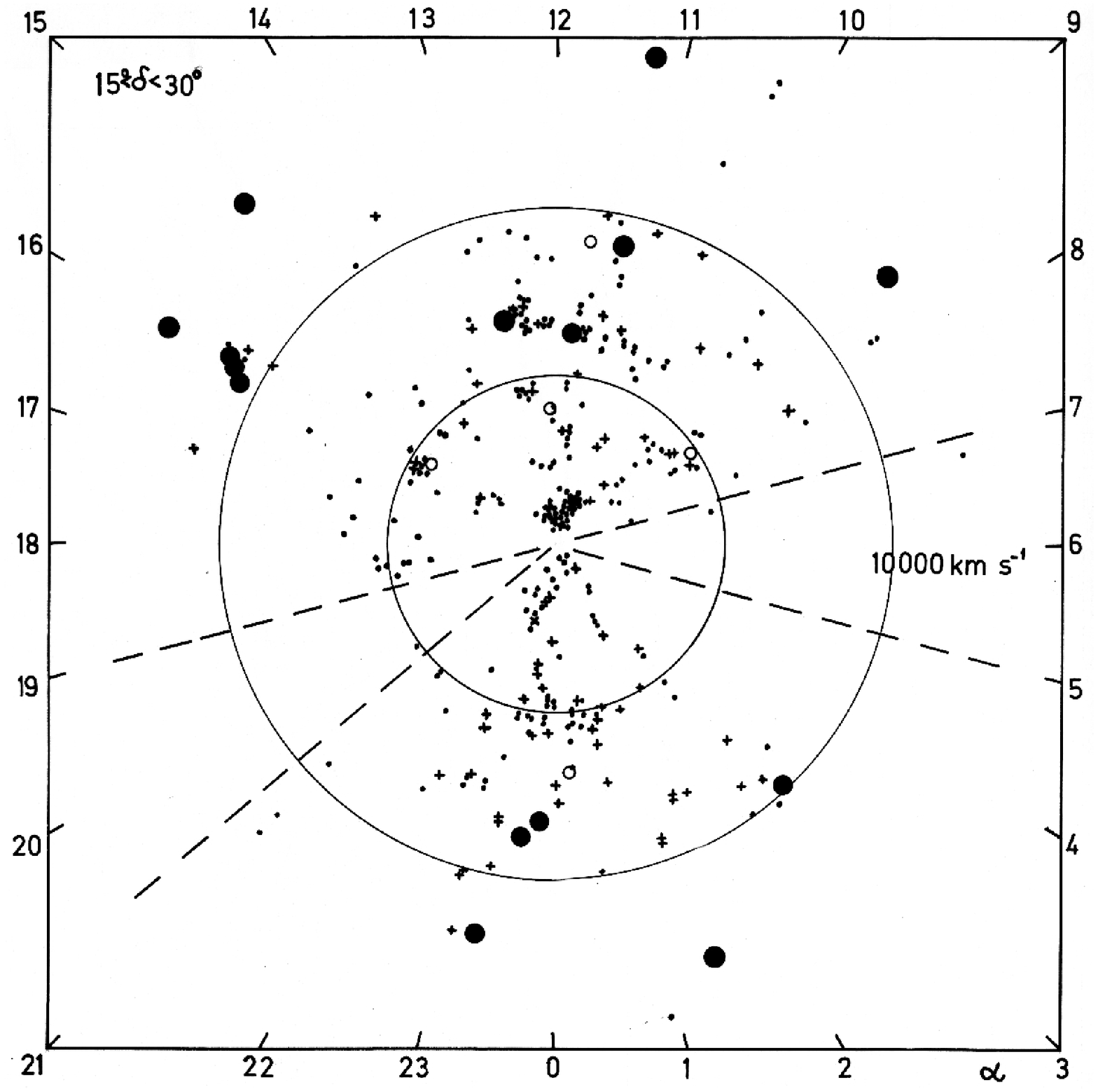}}
\vspace{3mm} 
\resizebox{0.48\textwidth}{!}{\includegraphics*{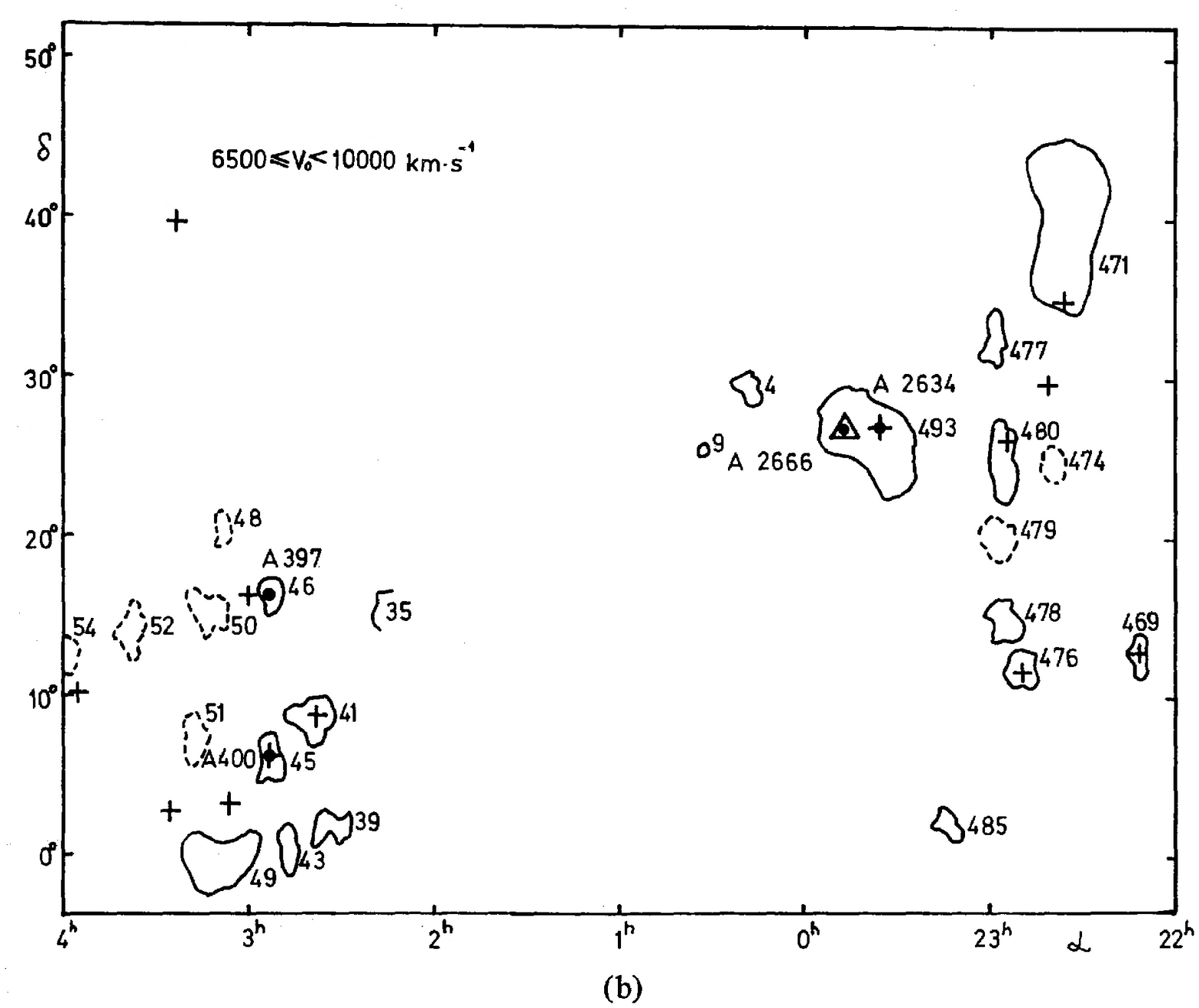}}\\
\resizebox{0.40\textwidth}{!}{\includegraphics*{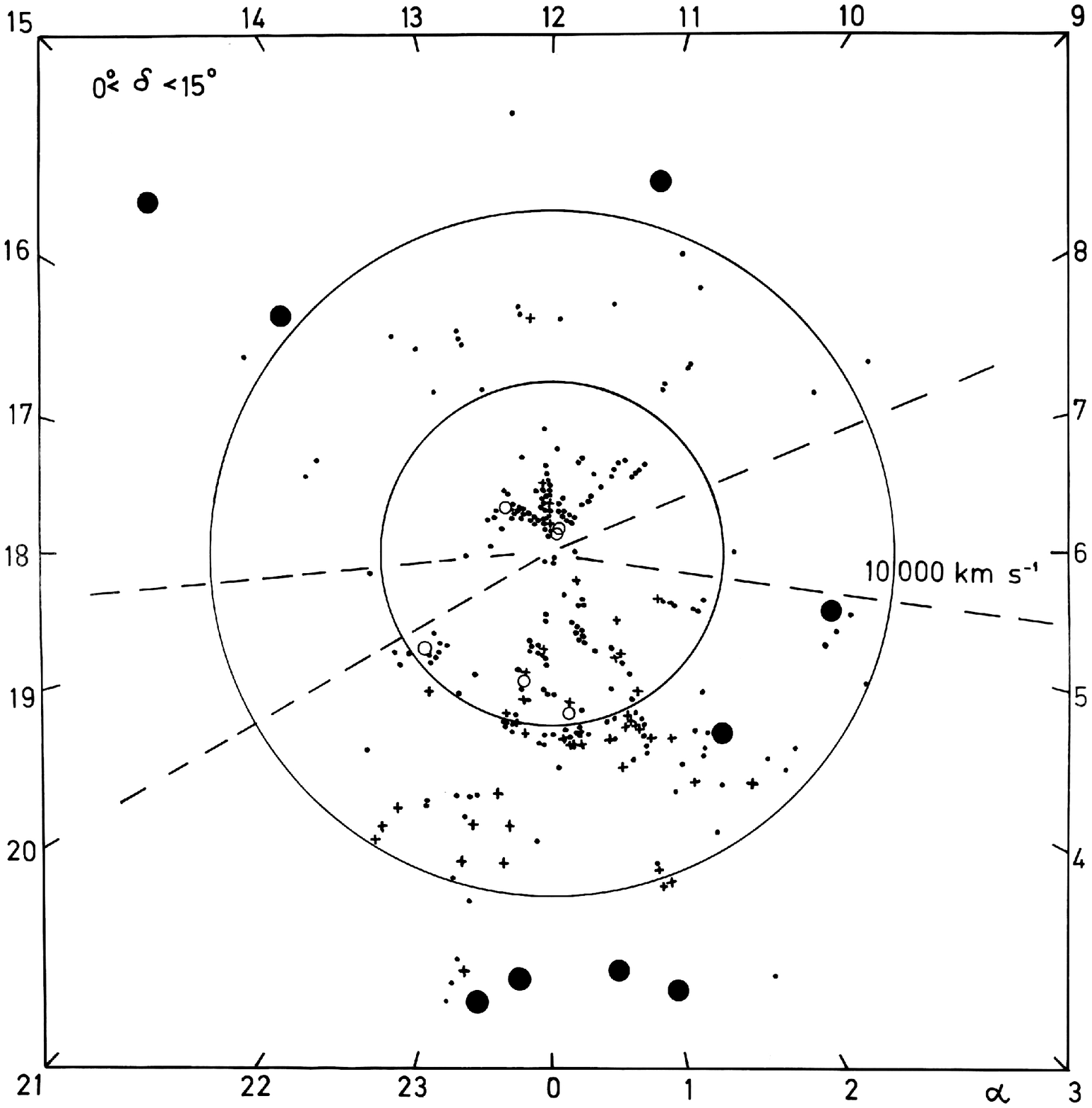}}
\resizebox{0.48\textwidth}{!}{\includegraphics*{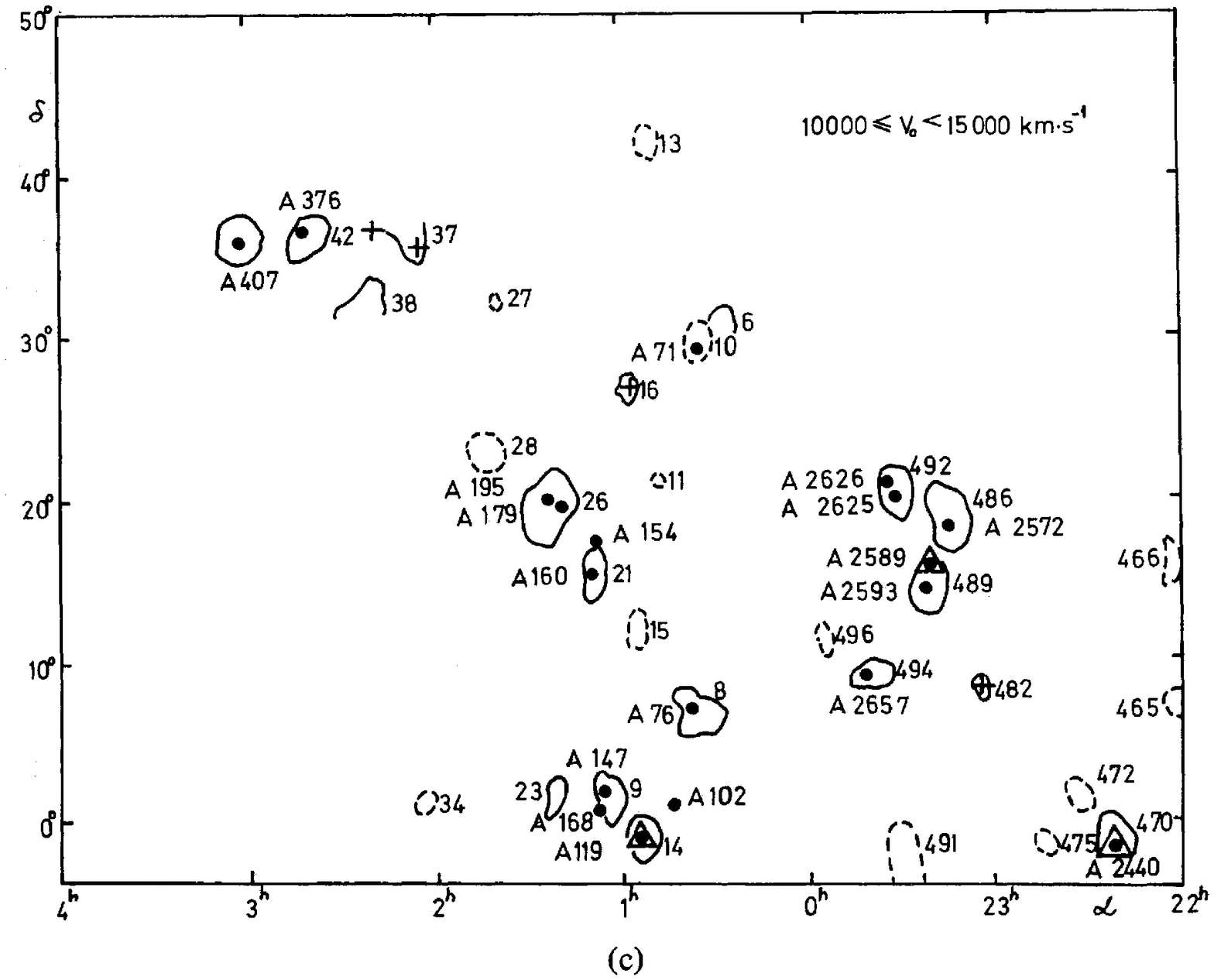}}
\caption{The distribution of galaxies and clusters, see text for explanations.  }
\label{fig:web}
\end{figure}

In our presentation we showed wedge diagrams of galaxies and clusters in
the Northern hemisphere, and galaxy and cluster plots in the Perseus
supercluster region, see Figure~\ref{fig:web}. In this Figure left
column shows wedge diagrams in three declination zones
(\cite{Joeveer:1978a}). Filled circles are for rich clusters of
galaxies, open circles --- groups, dots --- galaxies, crosses ---
Markarian galaxies.  In right panels we plot Abell clusters and contours
of Zwicky clusters in the Perseus area of sky at three distance
intervals  (\cite{Joeveer:1977py}, \cite{Einasto:1980}).  When combined these
pictures show the large-scale three-dimensional distribution of galaxies
and systems of galaxies.

Three-dimensional data showed the richness of the distribution of galaxies.
 Instead of random distribution of galaxies and clusters there
exists a complicated hierarchical network, which we called ``cellular
structure''.  Not only filaments (chains) of galaxies and clusters
were seen, but it was clear that galaxy chains form bridges between
superclusters. Thus there exists an almost continuous network of
superclusters and filaments. Some chains are rich and consists of
clusters and groups of galaxies, as the main ridge of the
Perseus--Pisces supercluster. Filaments of galaxies across large voids
are poor and consist only of galaxies and poor Zwicky clusters.  In
short, the three-dimensional data imply that the structure of the
Universe is much richer than believed so far. Presently it is called
the cosmic web.

Our picture had some similarity with the simulation made for the
Zeldovich pancake scenario.  However, the conclusion  was made only on
the basis of a visual impression by the comparison of the model and
observed distributions of particles/galaxies. \cite{Zeldovich:1978} in
his talk emphasised the need  to compare observations with models using
quantitative methods. Thus we started with Zeldovich and his
collaborators a search of quantitative methods to investigate
properties of the distribution of galaxies.

\begin{figure}[ht]
\centering
\resizebox{0.432\textwidth}{!}{\includegraphics*{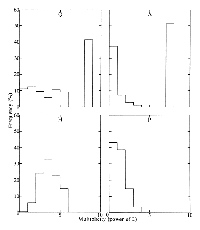}}
\resizebox{0.36\textwidth}{!}{\includegraphics*{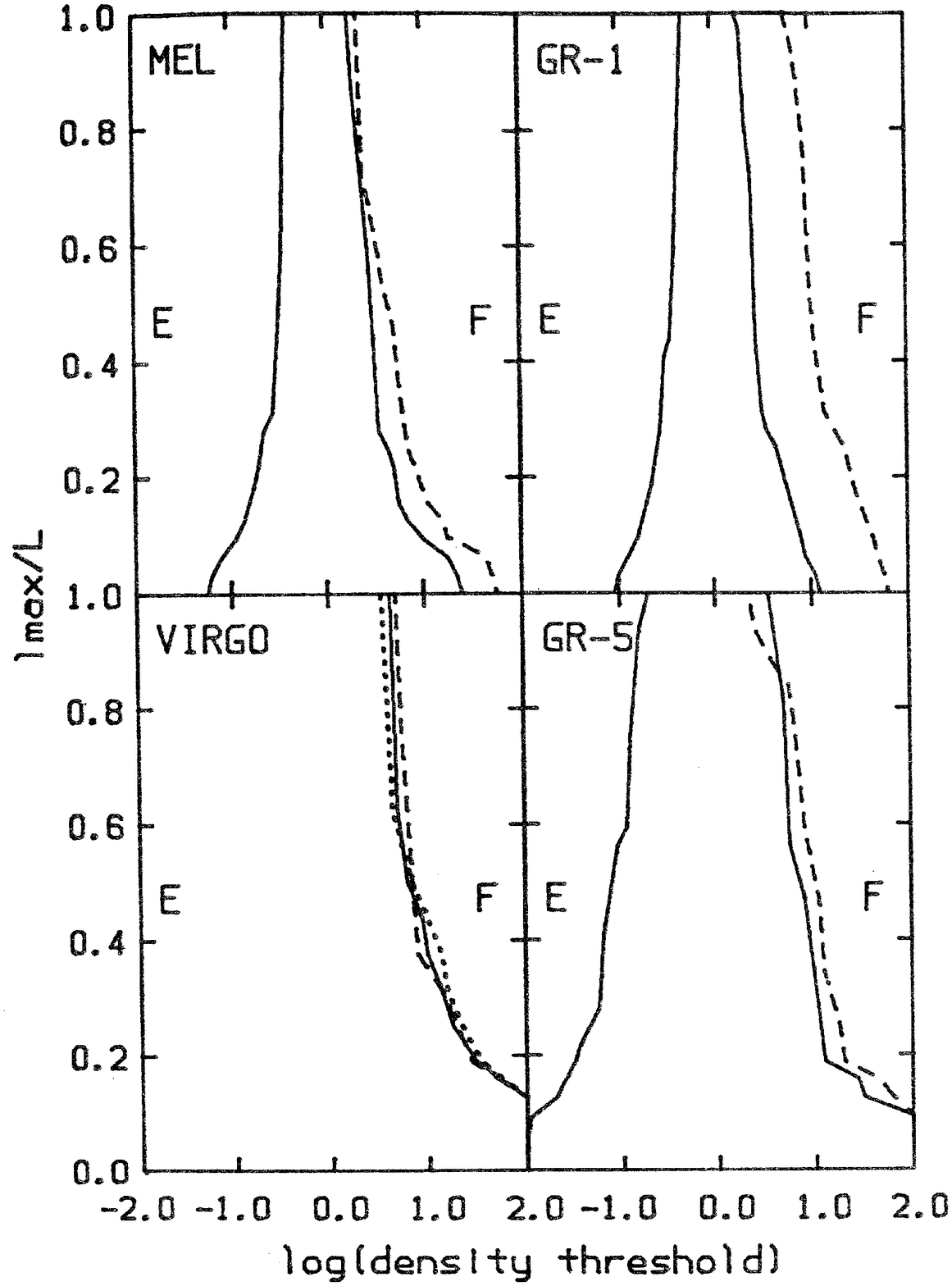}}\\
\caption{Left: the distribution of galaxies according to the
  multiplicity of the system. Multiplicity is expressed in powers of
  2.  Samples are designated: O --- observed, A --- adiabatic pancake
  model, H --- hierarchical clustering model, P --- Poisson model
  (\cite{Zeldovich:1982}).  Right: the length of the largest system
  (in units of the box size) versus the density threshold (in units of
  the mean density of the sample).  E is for low density (empty)
  regions, F for high density (filled) regions. MEL shows Melott CDM
  simulation, GR-1 and GR-5 Gramann LCDM simulation at expansion
  factors 1 and 5.2 (present epoch), and VIRGO the observed sample
  around the Virgo supercluster. In models solid lines indicate
  unbiased samples with all particles included, dashed lines show 
  biased samples, where particles in low-density regions have been
  removed (\cite{Einasto:1986ka}). Notice the similarity of the
  distribution of biased LCDM model at present epoch GR-5 with observations.}
\label{fig:tests}
\end{figure}

Our main results were published by \cite{Zeldovich:1982}. Here we used
the correlation function, the connectivity of systems of galaxies, the
length of the largest system calculated for various linking lengths, and
the multiplicity function of systems of galaxies.  Comparison was made
for a three-dimensional pancake model, hierarchical clustering model,
Poisson model, and observations (a volume limited sample of galaxies
including the Virgo supercluster).  These tests showed that in most
tests the pancake model is in good agreement with observations. In
contrast, the hierarchical clustering model is in conflict with all
tests, see the multiplicity test on left panel of Figure~\ref{fig:tests}.

However, some differences between the pancake model and observations
were evident. The most important difference is the lack of systems of
intermediate richness in the pancake model, observed in real galaxy
samples, as seen in the multiplicity test. As we understood soon, the
reason for this disagreement was the assumption that dark matter
consists of neutrinos.

\section{Astro-particle physics}

In early 1980's several important observational and theoretical
analyses were made which reinforced the need for a paradigm shift.  To
understand the nature of dark matter of key value were searches of
fluctuations of temperature fluctuations of CMB.  From theoretical
considerations it was clear that the temperature of CMB cannot be
constant, and the expected amplitude of fluctuations was $\delta T/T
\approx 10^{-3}$ (assuming the baryonic nature of the hot plasma
before recombination).  Fluctuations were searched with best radio
telescopes available, none was found and the upper limits were much
lower than the expected amplitude.

The most important theoretical development was the elaboration of the
inflation model of the early Universe by
\cite{Starobinsky:1980ys,Starobinsky:1982ly}, \cite{Guth:1981} and
\cite{Linde:1982kx}.  The inflation model was based on various
observational and theoretical considerations. One of the main
conclusions of the model is the prediction that the total
matter/energy density of the Universe must be exactly equal to the
critical cosmological density, $\Omega_{tot} =1$.

The observed density of baryonic matter in the Universe is about
$\Omega_b =0.05$, supported by primordial nucleosynthesis
considerations (mentioned already by \cite{Materne:1976} in the dark
matter discussion in Tbilisi, 1975).  Thus the only way to explain the
low level of CMB temperature fluctuations, and conclusions from the
inflation theory was to assume, that dark matter is
non-baryonic.  Non-baryonic matter is very weakly interacting with
radiation, and density fluctuations in the non-baryonic matter can
start to amplify already during the hot phase of the evolution of the
Universe.

These problems were discussed in April 1981 in a workshop in Tallinn,
where both particle physicists and astronomers attended.  A workshop
of similar topic was held in September-October in Vatican.  In both
workshops arguments were given for the non-baryonic nature of
dark matter.  These workshops mark the formation of a new area in
research --- astro-particle physics. 

The first natural candidate for the dark matter was neutrino, the only
known non-baryonic particle.  However, the problems with neutrinos as
dark matter candidate were soon realised, as discussed, among others,
by \cite{Zeldovich:1982}.  Thus astronomers and physicists started to
think what would be the alternatives. The main argument against
neutrinos was their very high speed, close to the speed of light,
which allowed to form only very massive cluster-sized systems.  To allow
the formation of smaller systems dark matter particles must have
higher mass and lover speed. So various hypothetical particles were
considered allowing the formation of systems of lower mass. Such
particles were commonly called Cold Dark Matter, in contrast to
neutrino-based Hot Dark Matter.

In 1983 Adrian Melott has made N-body simulations with density
perturbation spectra which corresponded to the hot dark matter as well
as to the cold dark matter scenario.  He visited Moscow and Tallinn to
discuss his results and to compare models with observations.  The
analysis was made jointly with Moscow and Tartu teams, and was
published by \cite{Melott:1983}.  Here we applied the same tests as
used by \cite{Zeldovich:1982}.  Our results showed that the CDM
model is in excellent agreement with all quantitative tests.  The 
paper ends with the conclusion, that the formation of the structure
starts with the flow of particles to form the filamentary web as in
the Zeldovich pancake model, but in the subsequent evolution
systems grow as in the hierarchical clustering scenario by Peebles.

The advantages of the CDM model were discussed in detail by
\cite{Blumenthal:1984}.  Now, finally the presence of dark matter was
accepted by leading theorists.  A very detailed series of N-body
simulations based on CDM and accepting a closed Universe with critical
density was made by the ``Gang of Four'' (\cite{Efstathiou:1985},
\cite{White:1987}).

In 1980's the attention of our cosmology team in Tartu was devoted to
quantitative study of the structure of the cosmic web using various
tests.  In these studies we used initially Melott CDM simulations to
compare observation with models of structure formation.  But we had
the need to have our own simulation to have full control of the model.
Enn Saar suggested to develop a model with cosmological $\Lambda$
term.  In this model the matter density was taken $\Omega_{m} = 0.2$,
as we have found from observations (\cite{Einasto:1974}).  The rest of
the matter/energy density is in the $\Lambda$ term, $\Omega_\Lambda =
1 - \Omega_{m} = 0.8$.  The simulation was made by our graduate student
Mirt Gramann.

Our LCDM model was used to investigate various properties of the
cosmic web. The first study was devoted to the topology of the cosmic web
by \cite{Einasto:1986ka}.  This study shows that the LCDM model fits
observational data even better than the standard CDM model with
critical density.  The need to use the LCDM model was discussed in detail
by \cite{Efstathiou:1990kh}.

Already in our first study of the cosmic web by \cite{Joeveer:1977py}
we had the question: Do galaxies form sheets between filaments as
expected in the pancake scenario, or are they formed only in high-density
regions as filaments at sheets crossing, and knots at filament
crossings.  The same question was asked also  by
\cite{Zeldovich:1982}, and studied on the basis of observational data
by \cite{Einasto:1980}.  The preliminary answer was --- there exists no
sheets of galaxies which isolate neighbouring low-density regions
between superclusters.  

\begin{figure}[t]
\centering
\resizebox{0.98\textwidth}{!}{\includegraphics*{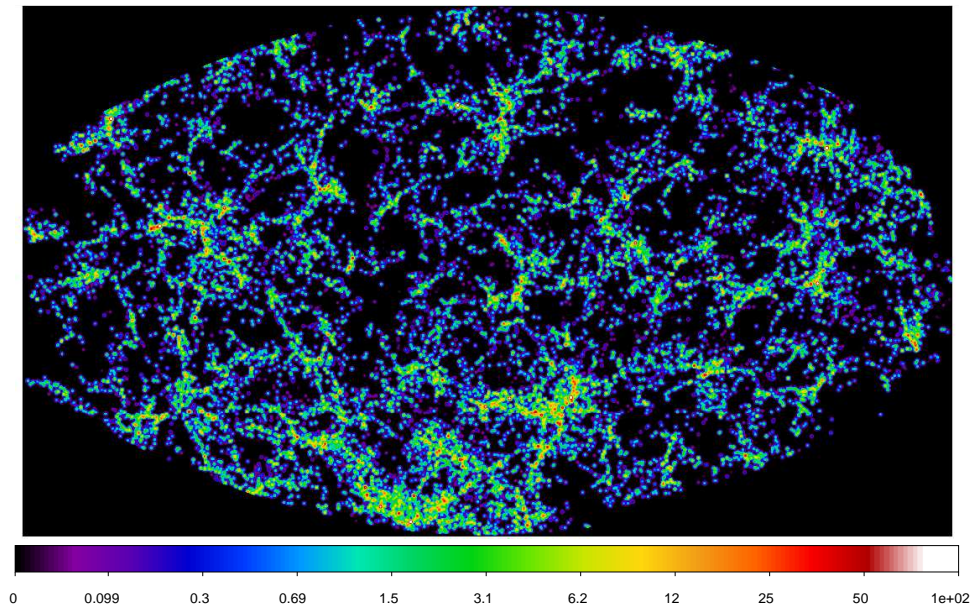}}
\caption{The luminosity density field of the SDSS in a spherical shell
  of 10~\Mpc\ thickness at a distance of 240~\Mpc.  The density scale is
  logarithmic, in units of the mean luminosity density for
  the whole Sloan Survey.  The rich complex in the lower area of the picture is
  part of the Sloan Great Wall; it consists of three very rich
  superclusters   (\cite{Suhhonenko:2011}).}
\label{fig:dr7}
\end{figure}

The more detailed study by \cite{Einasto:1986ka} showed, that the
topology of the web depends on the threshold density level applied to
separate low- and high-density regions in simulations.  At very low
threshold density sheets of particles isolate voids between rich
regions. However, in low-density regions there exists no conditions to
form galaxies.  The density of the collapsing gas must exceed a
threshold about 1.6 of the mean density to have during the Hubble time
the possibility to collapse, as shown by \cite{Press:1974fk}. If we
exclude particles from low-density regions (biased galaxy formation)
then voids (low-density regions in simulations) form just one large
connected region, both in real and model samples --- there are no sheets
isolating voids, see right panel of Figure~\ref{fig:tests}.  The length
of the largest system depends on the threshold level in all samples. 
Similar results were obtained by \cite{Gott:1986ly}.

Recent analysis of the luminosity density field shows the richness of
the cosmic web with rich and poor superclusters, galaxy filaments and
voids, see Figure~\ref{fig:dr7}.

An important aspect of the structure of the cosmic web is its fractal
character, as suggested by \cite{Mandelbrot:1982uq, Mandelbrot:1986fj}.
During a visit to NORDITA with Enn Saar we investigated the fractal
properties of the web in collaboration with Bernard Jones and Vicent
Martinez.  Our results showed that both observational and model samples
show multi-fractal properties (\cite{Jones:1988nu}).

One difficulty of the original pancake scenario  was the
shape of objects formed during the collapse.  It was assumed that
forming systems are flat pancake-like objects, whereas the dominant
features of the cosmic web are filaments (\cite{Joeveer:1978a},
\cite{Einasto:1980}).  This discrepancy was explained by
\cite{Bond:1996}, who showed that, due to tidal forces, in most cases
only essentially one-dimensional structures, i.e. filaments form.

\section{Summary}

The contribution of the Zeldovich team to modern cosmology is
impressive.  Zeldovich was very actively collaborating with other
groups, including our Tartu cosmology team.  Thank to close
collaboration with him and his team we jointly succeeded to
get interesting results on the nature of dark matter and the structure
of the cosmic web, and the connection between these two phenomena.

What impressed me most in the new cosmological paradigm is its beauty
--- the Universe is much richer than thought before.  The presence of
dark matter shows that the Nature of the Universe is richer: in addition
to known forms of matter it contains a new population, which is not
detected by particle physicists even today.  The Structure of the
Universe is also richer, instead of a random background of field
galaxies we see now the Cosmic Web with all its small and large details.

I thank all my collaborators in Tartu and Moscow for very fruitful
years of the search of properties of the Universe.

The present study was supported by ETAG project IUT26-2, and by the
European Structural Funds grant for the Centre of Excellence ``Dark
Matter in (Astro)particle Physics and Cosmology'' TK120.


\end{document}